\documentclass[12pt,pdflatex]{iopart}

\usepackage{iopams}
\usepackage{graphicx}
\usepackage{appendix}
\usepackage{wrapfig}

\begin{document}

\title[]{High-resolution imaging of ultracold fermions in microscopically 
tailored optical potentials}

\author{B Zimmermann$^{1,\ast}$\footnote[0]{$^{\ast}$ These authors 
contributed equally
to the presented work.}, T M\"{u}ller$^{1,\ast}$, J Meineke$^{1}$, T
Esslinger$^{1,\dag}$\footnote[0]{$^{\dag}$ Author to whom any
correspondence should be addressed.} and \\H Moritz$^{1,2}$ }

\address{$^1$Institute for Quantum Electronics, ETH Z\"{u}rich, 
H\"{o}nggerberg, CH-8093 Z\"{u}rich, Switzerland}
\address{$^2$Institut f\"{u}r Laser-Physik, Universit\"{a}t Hamburg, Luruper 
Chaussee 149, D-22761 Hamburg, Germany}

\ead{esslinger@phys.ethz.ch}


\begin{abstract}
We report on the local probing and preparation of an ultracold Fermi gas
on the length scale of one micrometer, i.e. of the order of the Fermi
wavelength. The essential tool of our experimental setup is a pair of
identical, high-resolution microscope objectives. One of the microscope
objectives allows local imaging of the trapped Fermi gas of $^{6}$Li atoms
with a maximum resolution of 660\,nm, while the other enables the
generation of arbitrary optical dipole potentials on the same length
scale. Employing a 2D acousto-optical deflector, we demonstrate the
formation of several trapping geometries including a tightly focussed
single optical dipole trap, a 4x4-site two-dimensional optical lattice and
a 8-site ring lattice configuration. Furthermore, we show the ability to
load and detect a small number of atoms in these trapping potentials. A
site separation of down to one micrometer in combination with the low mass
of $^6$Li results in tunneling rates which are sufficiently large for the
implementation of Hubbard-models with the designed geometries.

\end{abstract}

\maketitle

{\pagestyle{plain} \tableofcontents }

\section{Introduction}

Ultracold atomic Fermi gases provide a unique opportunity to create
strongly correlated many-body systems and to study intriguing phenomena,
such as the crossover between Bose-Einstein condensation and
Bardeen-Cooper-Schrieffer superfluidity~\cite{Varenna2006}. Experimental
access to this physics at a microscopic scale is now within reach as tools
are being developed to manipulate and observe quantum gases with high
spatial resolution~\cite{Nelson2007, Gericke2008, Bakr2009}. For a
fermionic system, a fundamental length scale is set by the Fermi
wavelength $\lambda_{F}$, which for instance determines the interparticle
distance and the scaling of density-density correlations. Typically,
$\lambda_{F}$ is of the order of one micrometer, a length scale which is
accessible by optical means.

Indeed, local optical probing has recently already allowed the in-situ
observation of suppressed density fluctuations in a degenerate Fermi
gas~\cite{Muller2010}. Extending this technique to local measurements of
density or spin fluctuations on the microscopic length scale of the
interatomic distance should give direct access to magnetic
properties~\cite{Recati2010,Sanner2011} and second-order, long-range
correlations~\cite{Yang1962}, such as the pair correlations characterizing
a superfluid Fermi gas. Thus, the ability to probe ultracold Fermi gases
on their natural length scale, the Fermi wavelength, will permit to gain
deeper insight into the mechanisms governing strongly correlated systems.

The manipulation of cold atoms on microscopic scales can be accomplished
with optical lattices~\cite{Bloch2005, Jaksch2005}. Ultracold fermions in
optical lattices constitute an almost ideal experimental realization of
the Hubbard model with highly tunable parameters~\cite{Esslinger2010}, and
very recently even single site resolution imaging has been achieved for
bosonic systems~\cite{Wurtz2009, Bakr2010, Sherson2010}. Yet, the concept
of optical lattices is by design restricted to the investigation of
periodic systems with a high degree of symmetry. Various approaches
towards more arbitrary, locally controllable geometries for optical
potentials have been successfully reported, for example double 
wells~\cite{Albiez2005,Boyer2006}, ring
traps~\cite{Heathcote2008,Henderson2009}, ring 
lattices~\cite{Henderson2009,Schnelle2008}, box 
potentials~\cite{Meyrath2005} and finite lattice 
patterns~\cite{Henderson2009,Schnelle2008,Dumke2002}. However, most 
realizations so far still lack the ability to shape optical
potentials on length scales comparable to the interatomic distance. Hence,
tunneling processes and dynamics are correspondingly very slow. To our
knowledge, experiments with fermions in microscopically tailored optical
potentials have not been reported yet.

In this work, we present the combination of both, the detection as well as
the preparation of ultracold fermionic samples on the microscopic length
scale of the Fermi wavelength. The main feature of our new experimental
apparatus is a pair of identical microscope objectives, each with an
optical resolution of 660\,nm at a wavelength of 671\,nm. One microscope
objective is part of a high-resolution imaging setup, while the other is
used for shaping versatile optical dipole potentials, which can be
tailored down to length scales below one micrometer. Using a two-axis
acousto-optical deflector, we demonstrate the site-by-site creation and
characterization of a tightly focussed single optical dipole trap, a
4x4-site two-dimensional square lattice and a 8-site ring lattice
structure. Moreover, we show the spatially resolved imaging of cold atoms
residing in the optically projected potential patterns.


\section{Preparation of a degenerate Fermi gas with high optical access}

We first outline the experimental concept of generating a quantum
degenerate Fermi gas of $^6$Li atoms, which is used as a reservoir for
loading atoms into the optical micro-potentials (see
sections~\ref{MicroPotential} and~\ref{AtomsInMicroPotential}). Details on
the experimental setup are given in ~\cite{Zimmermann2010}.

\begin{figure}
    \begin{center}
   \includegraphics[width=0.8\textwidth]{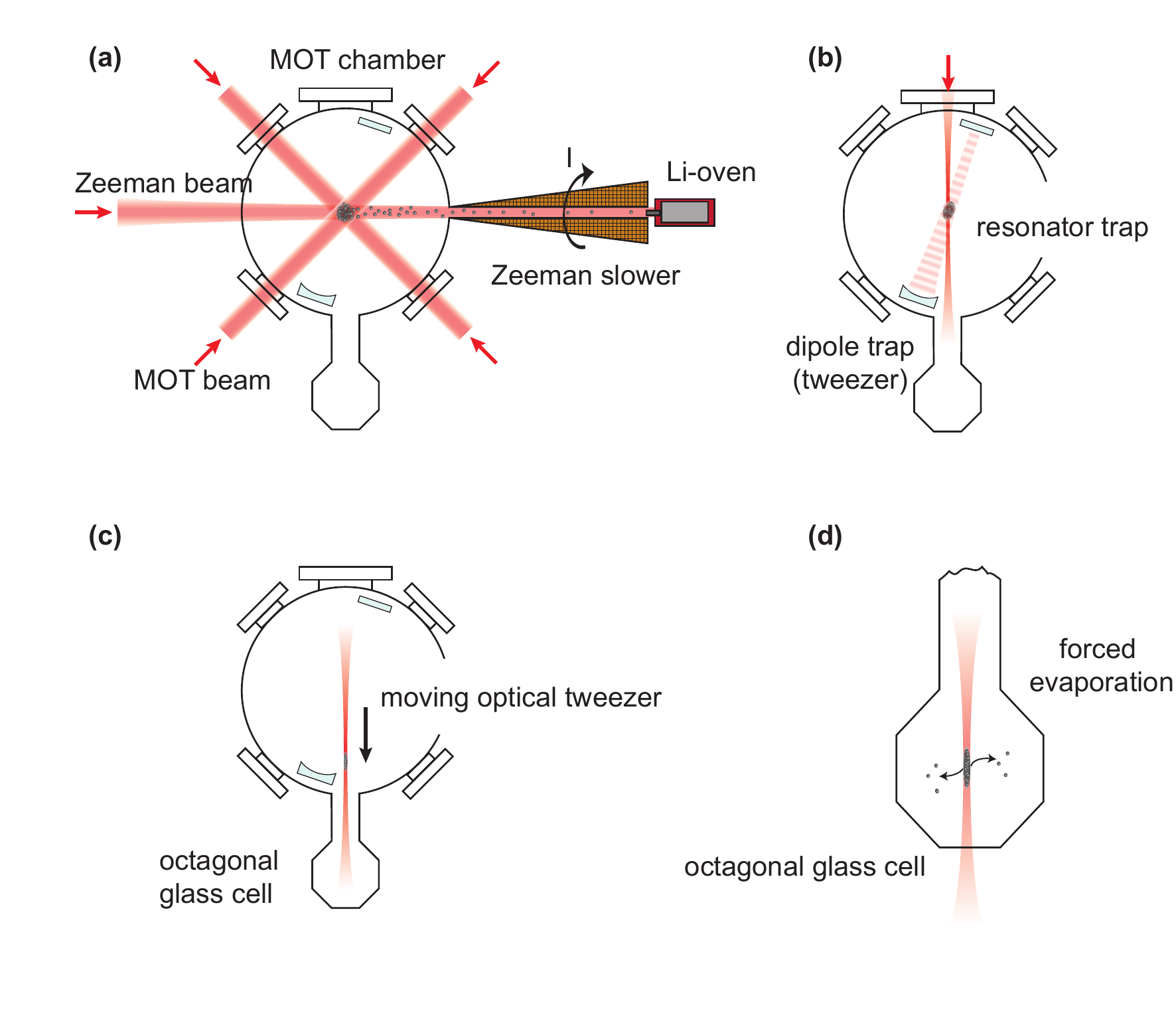}
    \end{center}
\caption[] {Experimental cycle: (a) $^{6}$Li atoms, emerging from the
oven, are decelerated by a Zeeman slower and captured in a magneto-optical
trap. (b) The trapped atoms are then transferred into a deep, large volume
optical dipole trap that is created using an optical resonator. In the
next step, the atoms are transferred into a tightly focused optical
tweezer. (c) By moving a lens, the atoms in the optical tweezer are
transported into the octagonal glass cell which offers exceptional optical
access. (d) Here, optical evaporation is performed by decreasing the power
in the optical tweezer, resulting in the formation of an ultracold Fermi
gas.}
    \label{fig:figure1}
\end{figure}

Our experimental apparatus consists of four sections which are
schematically depicted in figure~\ref{fig:figure1}: The oven chamber, the
Zeeman slower, the main ultra-high vacuum (UHV) chamber containing a high
finesse optical resonator, and the octagonal UHV glass cell attached to
the main vacuum chamber. The glass cell constitutes the final science
chamber offering high optical access for the microscopic detection and
manipulation of ultracold fermions. The preparation procedure of the
degenerate quantum gas - as illustrated in figure~\ref{fig:figure1}(a)-(d)
- has a duty cycle of approximately 12\,s and follows an all-optical
approach, similar to the methods applied in~\cite{Jochim2003}: $^6$Li
atoms emanating from the oven are decelerated by a Zeeman slower and
subsequently captured in a magneto-optical trap (MOT) at the center of the
main UHV chamber (see figure~\ref{fig:figure1}(a)). After 4\,s of loading,
the MOT typically contains 10$^9$ atoms at a temperature of about
$200\,\mu$K. In a second step, up to $9\cdot10^7$ atoms are transferred
into a large-volume standing-wave optical dipole trap realized by the
high-finesse optical resonator inside the main vacuum chamber (see
figure~\ref{fig:figure1}(b)). It consists of one flat and one curved
mirror (radius of curvature: 15\,cm) in hemi-spherical configuration, and
has a finesse of 10200, a resonator length of 14.975\,cm, a free spectral
range of 1\,GHz and a power enhancement factor of 1580. The resonator trap
is driven by far off-resonant laser light at a wavelength of 1064\,nm and
reaches a maximum trap depth of about $500\,\mu$K for a waist of
$500\,\mu$m (1/e$^2$-radius) at the MOT position. Here, the atomic sample,
which equally populates the two lowest hyperfine sub-states of $^6$Li, is
evaporatively pre-cooled. During the evaporation, we apply a magnetic
field of 300\,G to set the s-wave scattering length for the interstate
collisions of the two hyperfine sub-states to -300\,$a_0$, where $a_0$ is
the Bohr radius. The resonator trap serves as intermediate trapping
potential~\cite{Mosk2001}, which maximizes the particle transfer from the
MOT into the final trapping configuration, a tightly focused optical
tweezer (waist $w= 21\,\mu$m), also at a wavelength of 1064\,nm.
Typically, about $1.5\times10^6$ atoms in each sub-state are transferred
from the resonator trap into the running wave optical dipole trap, which
has a maximum initial trap depth of $\sim150\,\mu$K. By moving a lens over
a distance of 26.88\,cm~\cite{Gustavson2001}, the atoms in the optical
tweezer are then transported into the octagonal UHV glass cell (see
figure~\ref{fig:figure1}(c)). Here, forced evaporation is performed by
decreasing the power in the optical tweezer from 2\,W down to a few mW,
also at a magnetic field of 300\,G (see figure~\ref{fig:figure1}(d)). This
results in the formation of a quantum degenerate Fermi gas of
3$\times10^5$ atoms in each of the two lowest hyperfine sub-states at a
relative temperature of $\sim$0.3\,T$_F$, where $T_{F}$ is the Fermi
temperature. For the experiments shown in this paper, we typically prepare
the sample at this temperature.


\section{Optical system}

\begin{figure}[h]
\begin{center}
   \includegraphics[width=0.9
   \textwidth]{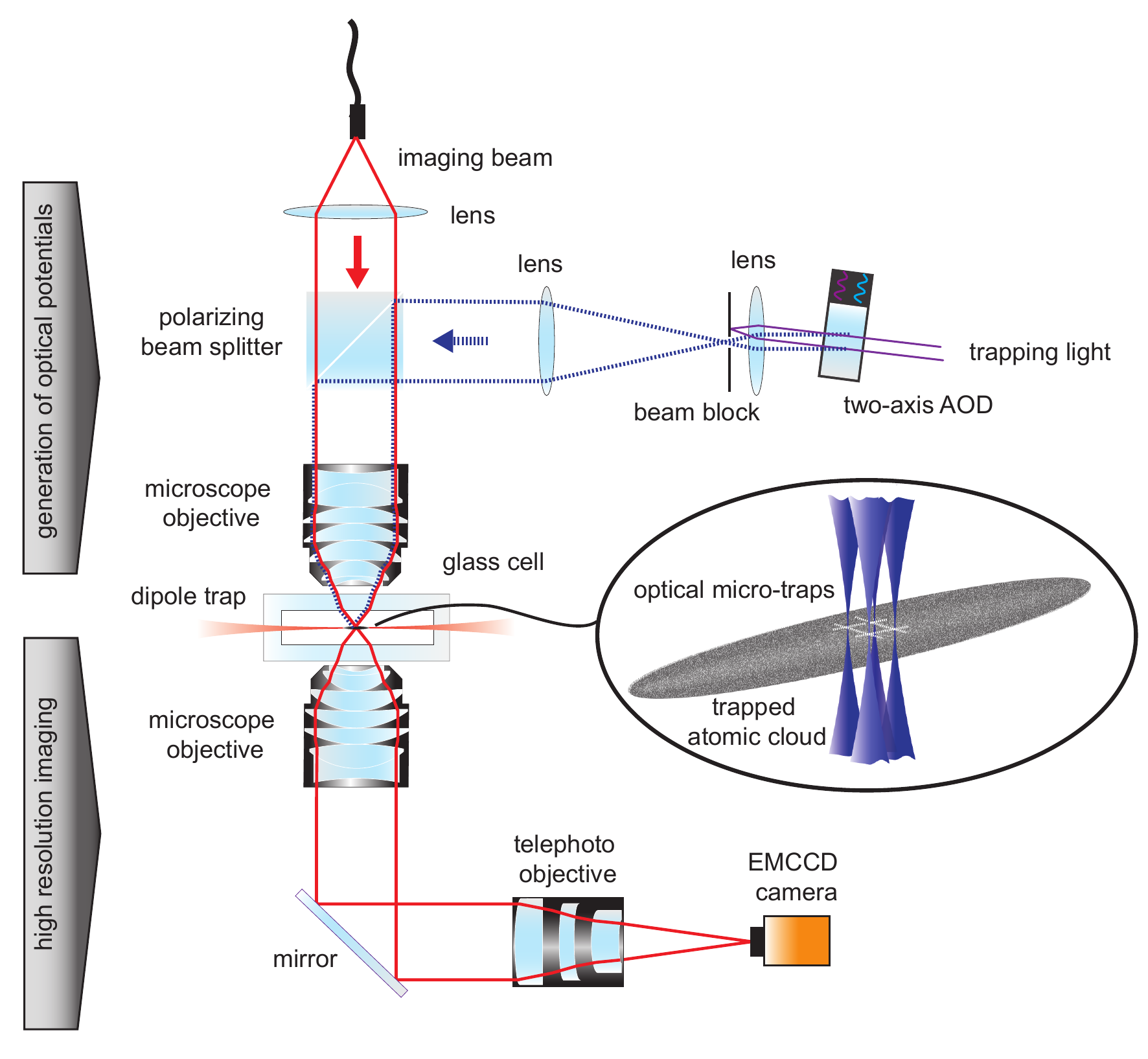}
    \end{center}
\caption[]{Two identical microscope objectives with high numerical
aperture constitute the heart of the high-resolution optical system: The
microscope objective below the glass cell and the telephoto objective
belong to the high-resolution imaging setup. Probe light, resonant to the
$|2S_{1/2}\rangle$ to $|2P_{3/2}\rangle$ transition along the D2-line of
$^6$Li, is collected by the microscope objective and imaged on an
electron-multiplying CCD camera (EMCCD). The second microscope objective
is part of the optical system for generating arbitrary optical
micro-potentials. A two-axis acousto-optical deflector generates several,
far off-resonant laser beams in a programmable way. Each of those beams is
focussed by the microscope objective, resulting in a controllable pattern
of multiple optical tweezers in the focal plane (see inset). The setup
allows for static as well as for time-averaged or dynamical optical
potentials. }
    \label{fig:figure2}
\end{figure}

The main feature of the experimental setup is the high-resolution optical
system as illustrated in figure~\ref{fig:figure2}, with two identical
microscope objectives centered along a common optical axis: one above the
glass cell, allowing for shaping optical dipole potentials down to length
scales below one micrometer; the other below the glass cell, used for
high-resolution imaging of the trapped quantum gas. We first describe the
technical details of the microscope objectives, before discussing their
applications for the microscopic imaging and manipulation of ultracold
fermions.

\subsection{High-resolution microscope objectives}

Each of the two identical, long-working-distance microscope objectives
(SPECIAL OPTICS, Inc.) is based on a system of 7 lenses as depicted in
figure~\ref{fig:figure2}. The infinite-conjugate configuration is designed
for three operating wavelengths at 532\,nm, 671\,nm and 770\,nm,
correcting aberrations of all three wavelengths. Furthermore, the
objectives are corrected for a view through the 4\,mm thick quartz window
of the glass cell. Each objective has an effective focal length of
f$\mathrm{_{eff}}$ = 18\,mm, and covers a numerical aperture of N.A. =
0.53. This yields a theoretical diffraction limit of 650\,nm (full-width
at half-maximum, FWHM) for an imaging wavelength of 671\,nm. All optical
surfaces are anti-reflection coated for the above wavelengths
(reflectivity R $< 0.75\%$) and additionally for 1064\,nm (R $< 2.0\%$).
The lenses are mounted in a non-magnetic and non-conducting housing made
of Ultem 3000 allowing for an undisturbed operation in the vicinity of the
magnetic coils around the glass cell.

We achieve the specified maximum optical performance at the diffraction
limit (RMS wavefront error $< 7\%$ of the wavelength) by an accurate
alignment of each microscope objective. The required constraints for the
absolute position accuracy only allow a displacement from the optical axis
by less than 1\,mm, a de-focussing smaller than 3\,$\mu$m, and a tilt of
the microscope axis with respect to the normal of the glass cell below
0.1$^\circ$. To meet these conditions the mounting of the microscope
objective includes a coarse and fine adjustment along five axes. The
tilting along two perpendicular axes can be adjusted with a goniometer
(NEWPORT, M-TTN80) without altering the height of the mount along the
central axis. This tilt platform has a centrical aperture with an inside
thread holding the microscope tube, which is made of machinable
glass-ceramic (MACOR). This permits a coarse focussing of the microscope
objective in axial direction (750\,$\mu$m per revolution). The transversal
coarse adjustment is performed by a two-axis translation stage (OWIS,
KT90), also with a centrical aperture. For the fine adjustment in
transversal and axial direction we use a compact, piezo-driven
three-dimensional translation stage (Piezo Jena, Tritor 102 SG EXT) with a
maximum translation distance of 100\,$\mu$m and a resolution of 3\,nm.
This device includes a strain gauge to measure the absolute position along
each axis. Via a digital proportional-integral controller based on a field
programmable gate array (National Instruments, NI 9264/9205/9401) the
error in position can be fed back to the piezo stage to actively stabilize
the transversal and axial position of the setup. The passive stability of
the goniometer is sufficient to keep the tilt angles within the
constraints.


\subsection{Imaging setup}

The high-resolution imaging system is sketched in the lower part of
figure~\ref{fig:figure2}, underneath the glass cell. It consists of three
components along the optical axis: The lower microscope objective, a
telephoto objective and an electron-multiplying CCD camera (EMCCD). The
custom-made telephoto objective (SPECIAL OPTICS, Inc.) is an
infinite-conjugate design with a focal length of approximately 1\,m. The
3-lens system is modeled for a joint operation with the microscope
objective at the diffraction limit and also corrects aberrations at
532\,nm, 671\,nm and 770\,nm. The lens surfaces of the telephoto objective
are AR-coated for the same wavelengths as the microscope objective. The
microscope objective is mounted 1.2\,mm below the glass cell, followed by
the telephoto objective at a distance of about 430\,mm. The EMCCD camera
is placed in the focal plane of the telephoto objective, about 450\,mm
behind its last lens. In total, the imaging system yields a magnification
factor of 54 and a field of view of $100\,\mathrm{\mu m}$
$\times100\,\mathrm{\mu m}$. In a calibration
measurement~\cite{Ottenstein2006}, the optical resolution was determined
to be about 660\,nm (FWHM) for a wavelength of 671\,nm. The depth of field
at full resolution is limited to $3.5\,\mathrm{\mu m}$, while the measured
chromatic axial focal shift was found to be well below 500\,nm for the
above wavelengths. For the detection of the imaging light we use a
back-illuminated EMCCD camera (ANDOR, iXon 897). Due to the magnification
factor of 54, each pixel ($16\,\mathrm{\mu m}\times16\,\mathrm{\mu m}$)
thus corresponds to $\sim300$\,nm$\times300$\,nm in the object plane.
Using the electron multiplier detected signals on the level of a few
incoming photons are amplified well above the read-out noise of the CCD
camera.

\subsection{Generation of optical micro-potentials}

We now describe the concept and setup for the creation of static and
time-averaged optical dipole potentials used to trap small samples of a
cold atomic gas. Similar to other work~\cite{Schnelle2008, Henderson2009,
Friedman2000, Onofrio2000}, we employ a two-axis acousto-optical deflector
(AOD) for the generation of structured optical trap configurations. The
AOD deflects and frequency shifts a red-detuned laser beam proportionally
to the frequency of the RF (radio-frequency) field fed into the AOD, while
its intensity can be controlled by the amount of RF power. The deflected
beam - enlarged by telescope optics - is then focussed by the upper
microscope objective onto the atomic cloud trapped in the large transport
dipole trap. Different deflection angles result in different positions of
the tweezer in the focal plane of the microscope objective as the latter
works as a Fourier transformer. The two-dimensional AOD is able to deflect
along two orthogonal axes and can be driven by multiple RF frequencies
along each axis at the same time. By this, versatile two-dimensional
multiple beam patterns can be created resulting from the convolution of
the beams deflected along the two perpendicular directions (see inset of
figure~\ref{fig:figure2}).

Our setup for generating multiple optical micro-potentials is sketched in
the upper part of figure~\ref{fig:figure2}, above the glass cell. As AOD
we use a two-axis deflector (IntraAction Corp., Model A2D-603AHF3A, center
frequency 60\,MHz, 3\,mm aperture) which incorporates a special acoustic
phased-array beam-steering design in order to maintain a uniform
diffraction efficiency (80$\%$) across the deflection bandwidth of 30\,MHz
(values for operation on only one axis). Additionally, a short access time
($\sim$276\,ns/mm beam diameter) of the AOD allows for the creation of
time-averaged potentials (see section~\ref{MicroPotential}). A collimated
laser beam (waist 1.2\,mm) at a wavelength of 767\,nm enters the AOD and
is subsequently deflected into the
(-1$^{\mathrm{st}}$/-1$^{\mathrm{st}})$-order for each applied RF
frequency. The deflected beams are then expanded by a two-lens telescope
to a maximum waist of 12.5\,mm. The microscope objective finally focusses
these collimated beams down to a diffraction limited spot size of about
730\,nm. With the opto-mechanical mounting for the microscope objective
described above, the resulting micro-trap pattern can be precisely aligned
and position-stabilized onto the atomic sample in the transport dipole
trap (see inset of figure~\ref{fig:figure2}). As RF-source we employ a
Universal-Software-Radio-Peripheral-2, which is controlled via the
gnuradio-software~\cite{USRP2,USRP2a}. It is capable of  generating
arbitrary wave forms with a bandwidth of 25\,MHz around a central
frequency (60\,MHz).

\section{Optical micro-potentials\label{MicroPotential}}

The configuration of our optical system with the two identical microscopes
allows us to directly monitor the optical potential landscape: the
high-resolution imaging setup below the glass cell accurately maps the
position, dimensions and intensity distribution of the trapping light
pattern shaped by the optical setup above the glass cell. Using this
information, we are able to characterize the trap geometry in terms of
waists, spacings, trap depths and trap frequencies. Moving the imaging
setup out of the focus of the upper microscope also provides insight into
trap parameters along the beam propagation direction.

\subsection{Single spot micro-trap}

\begin{figure}[h]
\begin{center}
   \includegraphics[width=1.0
   \textwidth]{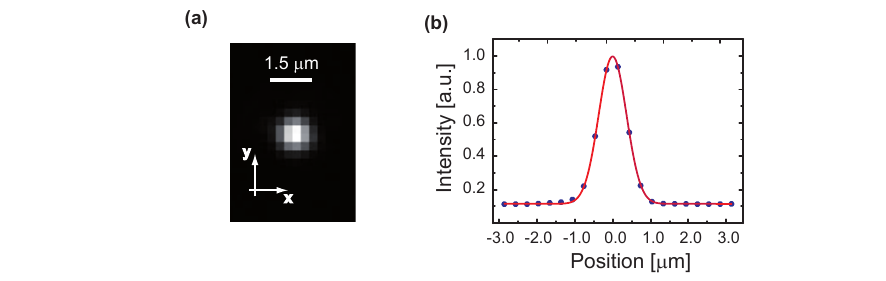}
    \end{center}
\caption[]{(a) High-resolution image of a single-spot dipole trap in its
focal plane (wavelength 767\,nm), illustrating the capability of the
optical system to map and characterize the optical potential landscape.
For the imaging, a bandpass filter for 671\,nm light in front of the CCD
camera is removed which normally blocks the trapping light in the
infra-red (767\,nm) when only the atoms trapped in the optical potential
are to be imaged. (b) Intensity profile of the single spot along the
$x$-axis. A Gaussian fit to the two-dimensional intensity distribution
yields the corresponding waists of the micro-trap.}
    \label{fig:figure3}
\end{figure}

Figure~\ref{fig:figure3}(a) shows the focal, two-dimensional intensity
distribution of a single spot created by the upper microscope as it is
imaged with high resolution by the lower microscope onto the EMCCD camera.
Fitting a Gaussian function to the measured intensity profiles along the
x- and y-axis yields a spot size with a waist (1/e$^2$-radius) of $w_{x} =
734$\,nm , and $w_{y} = 726$\,nm respectively (see
figure~\ref{fig:figure3}(b)). By moving the imaging system along the beam
propagation direction we measure the longitudinal intensity profile which
allows us to extract the Rayleigh length of the micro-trap. For the given
example in figure~\ref{fig:figure3}(a) the Rayleigh length is measured to
be $\sim2.1\,\mathrm{\mu m}$. With these parameters and a light power of
0.1\,mW, an optical dipole trap with a calculated depth of
$18.6\,\mathrm{\mu K}$ for $^6$Li atoms is created. The corresponding
trapping frequencies in radial direction are $\omega_{x} =
2\pi\cdot$69.5\,kHz and $\omega_{y} = 2\pi\cdot$70.2\,kHz, and $\omega_{z}
= 2\pi\cdot$16.5\,kHz along the axial confinement. A Gaussian trap is well
approximated by a harmonic potential up to $20\%$ of the trap depth,
whereas the total number of states in a harmonic trapping potential at
zero temperature with energy less than $\varepsilon$ is given by
$G(\varepsilon) =\frac {1}{6}
\frac{\varepsilon^3}{\hbar^3\omega_{x}\omega_{y}\omega_{z}}$. According to
this, for 0.1\,mW light power the single-spot dipole trap only offers
approximately one available state up to an energy level of $20\%$ of its
trap depth. However, when we load atoms into the micro-potential (see
section~\ref{AtomsInMicroPotential}), nearly all energy levels up to the
edge of the trap depth are populated. For the given parameters, an
interpolating expression~\cite{Simon2010} for the number of single
particle eigenstates of a Gaussian trap yields about 700 available states
up to  $99\%$ of the trap depth.

\subsection{Multiple spot micro-traps}

Figure~\ref{fig:figure4} and figure~\ref{fig:figure5} illustrate a
selection of possible multiple-spot potential patterns realized with our
setup. The images are again direct maps of the potentials in their focal
plane, imaged onto the EMCCD camera.

\subsubsection*{Two-dimensional finite optical lattice}

\begin{figure}[h]
\begin{center}
   \includegraphics[width=0.8
   \textwidth]{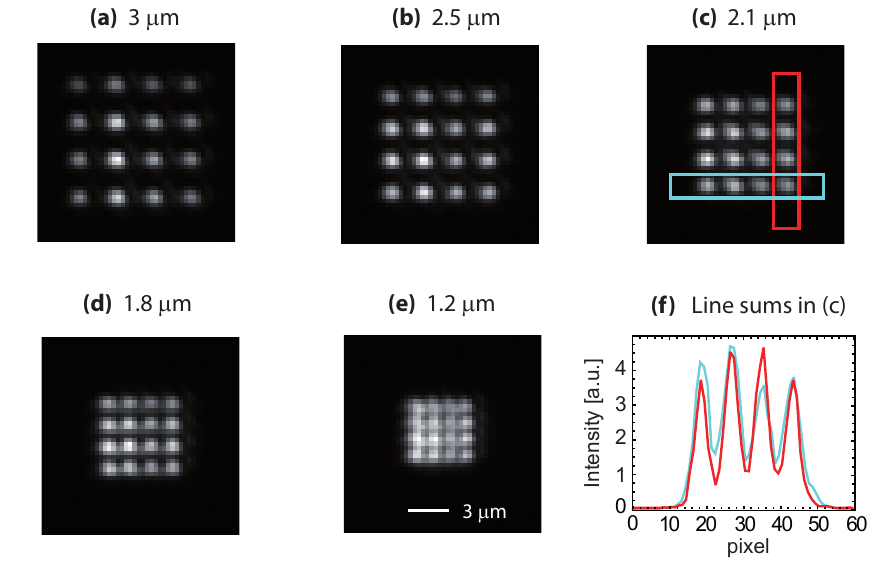}
    \end{center}
\caption[]{High-resolution images of a 4x4 site 2D lattice configuration
with different lattice spacings: (a) $3\,\mathrm{\mu m}$, (b)
$2.5\,\mathrm{\mu m}$, (c) $2.1\,\mathrm{\mu m}$, (d) $1.8\,\mathrm{\mu
m}$, and (e) $1.2\,\mathrm{\mu m}$. Each lattice site has a Gaussian spot 
size of approximately 800\,nm (1/e$^2$-radius). This spot size is slightly 
larger than for the single spot shown in figure~\ref{fig:figure3} since 
here, as well as for the ring lattice shown in figure~\ref{fig:figure5}, the 
numerical aperture of the upper microscope was reduced facilitating the 
alignment for this first demonstration. (f) Vertical and horizontal line sum 
profiles of the selected region in (c), marked in blue and red, 
respectively. The partial inhomogeneity in the intensity of different 
lattice sites results from an inhomogeneous diffraction efficiency within 
the RF bandwidth of the two-axis AOD. This inhomogeneity can be minimized, 
since the RF power for each RF frequency can be individually controlled, 
independently for both axes.}
    \label{fig:figure4}
\end{figure}

Applying four RF frequencies to each axis of the 2D AOD at the same time
results in a static 4x4 beam diffraction pattern in the
-1$^{\mathrm{st}}$/-1$^{\mathrm{st}}$-diffraction order. In the focal
plane of the upper microscope, this pattern yields a square array of 4x4
dipole traps which forms a finite-size, homogenous 2D optical lattice
system. Figure~\ref{fig:figure4}(a) shows such a lattice configuration.
Here, the four applied RF frequencies are symmetrically arranged around
the center frequency of 58\,MHz, separated by 7\,MHz. In real space, this
corresponds to a lattice site separation of approximately $3\,\mathrm{\mu
m}$ in the focal plane. Smaller lattice spacings can be achieved with
smaller RF frequency separations (figure~\ref{fig:figure4}(b-e)). The
spatial resolution of the imaging systems allows us to resolve lattice
spacings down to one micrometer as can be seen in
figure~\ref{fig:figure4}(e). The spacing can be changed dynamically even
during one experimental cycle, thus enabling e.g. the tuning of tunneling
dynamics within one experiment. Increasing the number of applied RF
frequencies easily enlarges this two-dimensional lattice pattern to a
maximum of 8x8 sites, limited by the finite RF deflection bandwidth of the
2D AOD. For the given configuration in figure~\ref{fig:figure4}(e), with a
lattice spacing of $1.2\,\mathrm{\mu m}$ and $10\,\mathrm{\mu W}$ light
power per lattice site, we estimate a tunneling rate of about 800\,Hz for
$^6$Li atoms populating the lowest Bloch band. These substantial tunneling
rates in combination with the possibility to tune the interparticle
interactions via Feshbach resonances thus provide a prospect for the
realization of Hubbard-model like physics in a finite-size, homogenous
lattice system.

\subsubsection*{Ring lattice}

\begin{figure}[h]
\begin{center}
   \includegraphics[width=1.0
   \textwidth]{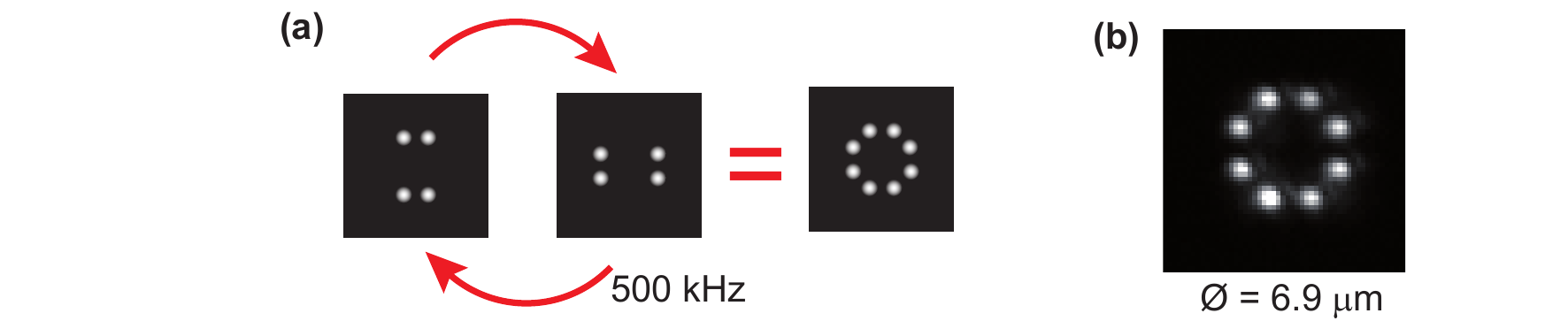}
    \end{center}
\caption[]{(a) Simulation data illustrating the generation of a
time-averaged ring lattice: two 2x2 rectangular square lattices - arranged
perpendicular to each other around a common symmetry axis - are
alternately projected onto the atoms. The switching frequency between the
two configurations is 500\,kHz, well above the trapping frequency of each
individual lattice site. (b) Real data image showing the light intensity
distribution of the resulting 8-site ring lattice with a diameter of
6.9$\,\mathrm{\mu m}$.}
    \label{fig:figure5}
\end{figure}

Apart from static potentials, our setup also allows us to generate
time-averaged optical potentials by alternately projecting different
trapping geometries onto the atoms. For this, the switching rate between
the different trap configurations has to exceed the corresponding trapping
frequency significantly in order to display a static trapping potential
for the atoms. In figure~\ref{fig:figure5} we give an example of such a
time-averaged optical potential. Here, we switch periodically with a
frequency of 500\,kHz between two different rectangular 2x2 lattice
configurations. This results in a 8-site ring lattice structure as
schematically sketched in figure~\ref{fig:figure5}(a).
Figure~\ref{fig:figure5}(b) presents the realization of this ring lattice
with our setup, showing the corresponding light intensity distribution in
the focal plane. In the given case, the ring diameter measures
$\sim6.9\,\mathrm{\mu m}$ with a Gaussian spot size of approximately 800\,nm 
(1/e$^2$-radius) for each lattice site. For the ring lattice the site 
separation can also be controlled arbitrarily down to about one micrometer 
as demonstrated for the square lattice configuration in the previous 
section.

\section{Atoms in micro-potentials\label{AtomsInMicroPotential}}

\begin{figure}[h]
\begin{center}
   \includegraphics[width=0.95
   \textwidth]{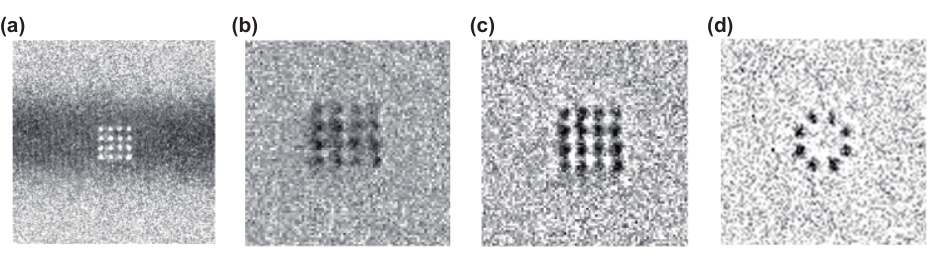}
    \end{center}
\caption[]{(a) 4x4-site 2D optical lattice superimposed to a fermionic
atom cloud trapped in the transport dipole trap. For this image, the
bandpass filter for 671\,nm in front of the CCD camera was removed,
allowing to image both, the trapped atom cloud with resonant absorption
imaging at 671\,nm and the focal light intensity distribution of the 4x4
lattice at 767\,nm. (b) Resonant absorption image of atoms trapped in the
lattice structure after a preparatory resonant light pulse was applied
which removed atoms in the shallow trapping regions. (c) Off-resonant
imaging of the same system, red-detuned by $\sim 10\Gamma$ with respect to
the upper hyperfine ground state of $^6$Li (illumination time: $12
\,\mathrm{\mu s}$). (d) Equivalent red-detuned imaging of atoms trapped in
the 8-site ring lattice pattern given in figure~\ref{fig:figure5}. All
images are divided by a reference image taken without atoms being
present.}
    \label{fig:figure6}
\end{figure}

The micro-trap patterns are filled from the reservoir of cold atoms
captured in the optical dipole trap which is used for the transport and
the final evaporative cooling in the glass cell. For this, the transport
and the micro-traps are spatially superimposed as sketched in the inset of
figure~\ref{fig:figure2}. Illustrating this situation,
figure~\ref{fig:figure6}(a) shows an in-situ absorption image of the
center part of an atomic sample held in the transport dipole trap (black
region). On top, a 4x4 lattice is imprinted whose intensity distribution
is simultaneously imaged, here appearing as white dots. For the transfer
of atoms, we smoothly ramp up the power of the micro-trap potential to its
final value (about 100$\,\mathrm{\mu W}$ per site) in 200\,ms and then
switch off the transport trap rapidly.

Imaging atoms in the micro-traps is challenging as the individual tightly
focussed laser beams spatially overlap at a certain distance along the
axial direction, depending on their separation. Moreover, the atoms
typically populate energy levels up the edge of the trap right after the
transfer. Therefore, by resonant absorption imaging along the microscope
axis the individual atomic samples in different trapping potentials appear
as a continuous shadow cast on the CCD camera. However, the filling of the
micro-traps can be reduced by sending resonant light onto the atoms prior
to the imaging. For this purpose, we apply a resonant light pulse of
$4\,\mathrm{\mu s}$ at one tenth of the $^6$Li saturation intensity
(2.54\,mW/cm$^2$) onto the sample. This removes the atoms in trap regions
of shallow potential depth and only leaves atoms in the tightly confining
center. By a second resonant light pulse ($24\,\mathrm{\mu s}$) at twice
the saturation intensity, the remaining atoms are then imaged through the
microscope setup onto the CCD camera. Figure~\ref{fig:figure6}(b) shows
the in-situ absorption image of the remaining atoms trapped in a 4x4
two-dimensional square lattice with a site separation of $2.5\,\mathrm{\mu
m}$.

Alternatively, due to the high optical density, the atoms can be imaged
dispersively with off-resonant laser light, which reveals the atoms
captured in the center of each micro-trap. For this, we red-detune the
imaging light along the $|2S_{1/2}\rangle$ to $|2P_{3/2}\rangle$
transition by about $10\Gamma$  with respect to the second lowest
hyperfine sub-state of $^6$Li. Here, $\Gamma$ = 5.9\,MHz is the natural
line width of the D2 transition of $^6$Li. The resulting off-resonant
dispersive image can be seen in figure~\ref{fig:figure6}(c) where the
atoms in the different trap centers of the 4x4 lattice appear as well
separated dark spots. In this case, a preparatory resonant light pulse in
advance is not required. We applied the same off-resonant imaging
technique to atoms trapped in the 8-site ring lattice structure
(figure~\ref{fig:figure5}(b)), see figure~\ref{fig:figure6}(d).

The number of trapped atoms can be estimated in a time of flight (TOF)
measurement: Releasing the sample from the confining optical potential the
expanding cloud is imaged after a certain TOF by means of resonant
absorption imaging through the microscope. From a fit to the detected
density profile we determine the total atom number. For the situation in
figure~\ref{fig:figure6}(c), the fit yields an upper limit of 300 atoms
per lattice site. We also measured the lifetime of the trapped sample in a
separate experiment. For the static potential we observe a two-stage loss
process of the trapped atoms. In an initial fast decay on a short
timescale ($\sim$100\,ms), the atom number is approximately reduced by a
factor of 2. Subsequently, the population decays exponentially on a longer
timescale, yielding a lifetime of 800\,ms. For the time-averaged ring
lattice, we find shorter lifetimes of about 100\,ms. A recent measurement of 
inelastic collisions in a two-component Fermi gas prepared in the strongly 
interacting BEC-BCS crossover~\cite{Du2009} showed that two- and three-body 
collisions give rise to particle losses on a time scale well above one 
second. Our samples have a comparable density, yet are weakly interacting 
($a= -300\,a_0$) and hence the effect of three-body collisions on the 
observed lifetime are expected to be negligible. In addition, we can exclude 
light scattering to affect the lifetime of the trapped sample. For the given 
trap parameters, the photon scattering rate  is 0.25\,Hz per atom. Most 
likely, the particle loss is caused by free evaporation from the initially 
completely filled micro-traps. Moreover, the intensity of the micro-trap was 
not actively stabilized for the presented measurement, possibly causing 
spilling of particles from the trap due to fluctuations in the trap depth. 
The even faster loss rates observed for the ring lattice potential are 
probably induced by heating of the sample due to the fact that this 
particular trapping configuration results from a time averaged projection of 
two rectangular lattice patterns. Further investigation is needed to fully 
understand the observed lifetimes.

\section{Conclusion and outlook}

We have presented an experimental setup with two high-resolution
microscope objectives that allows us to optically probe and prepare an
ultracold Fermi gas on the microscopic length scale of the Fermi
wavelength. Employing a 2D acousto-optical deflector, we have demonstrated
the site-by-site generation of a finite two-dimensional square lattice and
a 8-site ring lattice configuration. Moreover, we have shown the capability 
to load small numbers of atoms into these optical micro-potentials and to 
detect them with single-site resolution.

An immediate, although non-trivial extension of this work would be the 
measurement of the temperature in the micro-traps. Currently, the measured 
particle numbers and the known trap geometry should lead to a Fermi 
temperature close to the trap depth which gives us confidence in the 
assumption that the trapped sample might be quantum degenerate.

The expected substantial tunneling amplitudes for $^6$Li atoms in 
combination with the possibility to tune the inter-particle interactions via 
Feshbach resonances promises a possible realization of Hubbard-model like 
physics beyond the standard optical lattice approach with interfering laser 
beams. While the site-by-site creation of lattice sites has the advantage of 
the intrinsic absence of additional external confinement, it also offers the 
ability to generate lattice systems of low symmetry or systems with inherent 
defects. In addition, our setup holds the potential for a single-site 
addressability allowing the individual manipulation of atoms in different 
trap spots~\cite{Wurtz2009,Dumke2002,Scheunemann2000,Karski2010}.

\section*{Acknowledgments}
\addcontentsline{toc}{section}{Acknowledgments}

We acknowledge J.-P. Brantut and D. Stadler for experimental assistance
and valuable discussions. We also thank the group of M. Oberthaler
(Heidelberg) for the loan of a test target. This work was supported by the
ERC advanced grant SQMS, the FP7 FET-open grant NameQuam and the NCCR
MaNEP. Mention of industrial brand names is for technical communication
only and does not constitute an endorsement of such products.



\end{document}